\documentclass[journal]{IEEEtran}
\usepackage{eurosym}
\usepackage{color}
\usepackage{refcount}
\usepackage[numbers,sort&compress]{natbib}
\usepackage{setspace}
\usepackage{flushend}
\usepackage[caption = false]{subfig}
\usepackage{float}
\usepackage[normalem]{ulem} % either use this (simple) or

\usepackage[pdftex]{graphicx}
\DeclareGraphicsExtensions{.pdf, .pdf.bb}
\ifCLASSINFOpdf

\else
\fi
\usepackage[cmex10]{amsmath}
\def\mathclap#1{\text{\hbox to 0pt{\hss$\mathsurround=0pt#1$\hss}}}

\usepackage{fixltx2e}
\usepackage{amssymb}
\usepackage{varioref}
\usepackage{slashbox}

\usepackage{amsmath}% http://ctan.org/pkg/amsmath
\usepackage{algorithm,algcompatible}% http://ctan.org/pkg/{algorithms,algorithmicx}
\algnewcommand\algorithmicreturn{\textbf{return}}
\algnewcommand\RETURN{\algorithmicreturn}
\algnewcommand\algorithmicprocedure{\textbf{procedure}}
\algnewcommand\PROCEDURE{\item[\algorithmicprocedure]}%
\algnewcommand\algorithmicendprocedure{\textbf{end procedure}}
\algnewcommand\ENDPROCEDURE{\item[\algorithmicendprocedure]}%
\algnewcommand{\algvar}[1]{{\text{\ttfamily\detokenize{#1}}}}
\algnewcommand{\algarg}[1]{{\text{\ttfamily\itshape\detokenize{#1}}}}
\algnewcommand{\algproc}[1]{{\text{\ttfamily\detokenize{#1}}}}
\algnewcommand{\algassign}{\leftarrow}

\usepackage{makeidx}
\makeindex

\begin{document}

%
% paper title
% can use linebreaks \\ within to get better formatting as desired
% Do not put math or special symbols in the title.
\title{  A Machine Learning-based Approach to Build Zero False-Positive IPSs for Industrial IoT and CPS with a Case Study on Power Grids Security }
%Cyber Attack Prevention through  Zero False-Positive Rule-Set Generation for Industrial Systems Firewalls \\Prevention of Cyber Attacks on Industrial Systems through  Zero False-Positive   Rule-Set Generation for     Firewalls\\
%
% author names and IEEE memberships
% note positions of commas and nonbreaking spaces ( ~ ) LaTeX will not break
% a structure at a ~ so this keeps an author's name from being broken across
% two lines.
% use \thanks{} to gain access to the first footnote area
% a separate \thanks must be used for each paragraph as LaTeX2e's \thanks
% was not built to handle multiple paragraphs
%
\author{Mohammad~Sayad~Haghighi,~\IEEEmembership{Senior Member,~IEEE,}
        Faezeh Farivar,~\IEEEmembership{Member,~IEEE}\\% <-this % stops a space
\thanks{Manuscript received Sep. 2019.}
\thanks{M. Sayad Haghighi (corresponding author) is with the School of Electrical and Computer Engineering, University of Tehran, Iran, Email: sayad@ieee.org. }\thanks{F. Farivar is with the Department of Mechatronics and Computer Engineering, Science and Research Branch, Islamic Azad University, Tehran, Iran, Email: F.Farivar@srbiau.ac.ir.} 
}
% note the % following the last \IEEEmembership and also \thanks - 
% these prevent an unwanted space from occurring between the last author name
% and the end of the author line. i.e., if you had this:
% 
% \author{....lastname \thanks{...} \thanks{...} }
%                     ^------------^------------^----Do not want these spaces!
%
% a space would be appended to the last name and could cause every name on that
% line to be shifted left slightly. This is one of those "LaTeX things". For
% instance, "\textbf{A} \textbf{B}" will typeset as "A B" not "AB". To get
% "AB" then you have to do: "\textbf{A}\textbf{B}"
% \thanks is no different in this regard, so shield the last } of each \thanks
% that ends a line with a % and do not let a space in before the next \thanks.
% Spaces after \IEEEmembership other than the last one are OK (and needed) as
% you are supposed to have spaces between the names. For what it is worth,
% this is a minor point as most people would not even notice if the said evil
% space somehow managed to creep in.

% The paper headers
\markboth{}%
{Shell \MakeLowercase{\textit{et al.}}: Bare Demo of IEEEtran.cls for Journals}
% The only time the second header will appear is for the odd numbered pages
% after the title page when using the twoside option.
% 
% *** Note that you probably will NOT want to include the author's ***
% *** name in the headers of peer review papers.                   ***
% You can use \ifCLASSOPTIONpeerreview for conditional compilation here if
% you desire.

% If you want to put a publisher's ID mark on the page you can do it like
% this:
%\IEEEpubid{0000--0000/00\$00.00~\copyright~2012 IEEE}
% Remember, if you use this you must call \IEEEpubidadjcol in the second
% column for its text to clear the IEEEpubid mark.

% use for special paper notices
%\IEEEspecialpapernotice{(Invited Paper)}

% make the title area
\maketitle

% As a general rule, do not put math, special symbols or citations
% in the abstract or keywords.
\begin{abstract}
Intrusion Prevention  Systems (IPS),  have long been the first layer of defense against malicious attacks. Most sensitive systems employ instances of them (e.g. Firewalls) to secure the network perimeter and filter out attacks or unwanted traffic.  A firewall, similar to classifiers, has a boundary to decide which traffic sample is normal and which one is not.  This boundary is defined by configuration and is managed by a set of rules  which occasionally might also filter normal  traffic by mistake. However, for some applications, any interruption of the normal operation is not tolerable e.g.  in power plants, water distribution systems,  gas or oil pipelines, etc. In this paper, we  design a learning firewall that receives labelled samples and configures itself automatically by writing preventive rules in a conservative way that avoids false alarms. We design a new family of classifiers, called $\mathfrak{z}$-classifiers, that unlike the traditional ones which merely target accuracy, rely on zero false-positive as the metric for decision making. First, we analytically show why naive modification of current classifiers like SVM does not yield acceptable results and then, propose a generic iterative   algorithm  to accomplish this goal. We  use the proposed classifier with CART at its heart to build a firewall for   a Power Grid Monitoring System. To further evaluate the  algorithm, we additionally test it on KDD CUP'99 dataset. The results confirm the effectiveness of our approach.

\end{abstract}

% Note that keywords are not normally used for peerreview papers.

% For peer review papers, you can put extra information on the cover
% page as needed:
% \ifCLASSOPTIONpeerreview
% \begin{center} \bfseries EDICS Category: 3-BBND \end{center}
% \fi
%
% For peerreview papers, this IEEEtran command inserts a page break and
% creates the second title. It will be ignored for other modes.
\IEEEpeerreviewmaketitle

\section{Introduction}
% The very first letter is a 2 line initial drop letter followed
% by the rest of the first word in caps.
% 
% form to use if the first word consists of a single letter:
% \IEEEPARstart{A}{demo} file is ....
% 
% form to use if you need the single drop letter followed by
% normal text (unknown if ever used by IEEE):
% \IEEEPARstart{A}{}demo file is ....
% 
% Some journals put the first two words in caps:
% \IEEEPARstart{T}{his demo} file is ....
% 
% Here we have the typical use of a "T" for an initial drop letter
% and "HIS" in caps to complete the first word.
\IEEEPARstart{I}{ntrusion} Prevention Systems (IPS)  are security systems that monitor and control the traffic coming in or going out of the network based on a set of pre-defined rules \cite{zhang2004intrusion}. 
They usually protect the perimeter of the networks and keep the insiders away from the malicious  attacks launched from outside. The most well known examples of IPSs are firewalls.

In delay sensitive applications, where continuity of operation is vital,   blocking a legitimate traffic  by mistake is not tolerable. This could be  disruption in e.g. industrial Cyber Physical Systems  (CPS) or governmental services. Some tried solving the issue by mutual authentication of interacting entities \cite{mohammadi2008robust}, however, the majority rely on IPSs. Classic intrusion detection mechanisms rely on classification of samples and divide them into benign and malicious \cite{ liao2013intrusion,sayad2019, farivar2019artificial}. The border is usually drawn by using a cost minimization algorithm. However, this does not guarantee a zero false alarm and occasionally legitimate traffic is filtered.

Currently, firewalls are mostly configured by experts manually and with the update of attack samples, they are hardly updated, since modifying the rule sets becomes very complex especially when you do not want to tamper with the normal traffic in organizations and industrial applications where business continuity is crucial \cite{toorchi2013markov, arabsorkhi2016conceptual}. Leaving all the decision makings to  Intrusion Detection Systems (IDS) is also problematic as IDSs are usually passive systems and have processing limitations which hinder the precise inspection of samples when they are overloaded  \cite{alshammari2007using,qiao2002network, joo2003neural, haghighi2010neighbor,  barchinezhad2020compensation, haghighi2019intelligent}. 

In this paper we aim to design a next generation learning firewall that receives labelled samples (e.g. from a supervised Intrusion Detection System (IDS)) and configures itself automatically by writing preventive rule sets in a conservative way that avoids false alarms. 
In other words, we assume   that malicious and normal samples are fed to an automated algorithm of writing preventing rules in the firewall so that it does not block any normal or benign traffic (zero false positive). This is vital in applications where normal operation interruption is not tolerable e.g. in governmental services or cyber-physical systems like power plants, water distribution systems, etc. 
We design a new family of classifiers that unlike the traditional ones which merely target accuracy, rely on zero-false-positive as the metric for decision making. We try to use iterative  methods to accomplish this. The cost paid is an increase in the false negative rate which we tend to minimize. 

We leave the grey area to the IDS as it may evolve and improve over the time. However, the learning firewall proactively prevents the damage compared to the passive approach of letting the traffic in and then decide.  Besides, it will reduce the load of the IDS  too so that it can then increase the complexity or depth of analysis for the remaining traffic. 
At the final stage,  the output of this new classifier is automatically translated into non-conflicting firewall rules. We conduct some experiments on real-world datasets obtained form industrial systems to show how the our approach works.  
\\
The rest of this paper is organized as follows: In Section~\ref{rel} we review the related work.  Section \ref{section:proposed_model} studies the problem of achieving zero false positive rates and then presents a novel algorithm to make it possible. {Section~\ref{section:results} tests the algorithm on a power grid monitoring system as well as KDD~CUP'99 datasets.  Finally, the paper is concluded in Section~\ref{section:conclusion}.

\section{Related Work\label{rel}}
%According to the contributions of this paper, we discuss the related work in the following three subsections.
%\subsection{Propagation Modelling in WSNs}
 Several  papers focused on increasing the accuracy, speed of traffic recognition or  rule enforcement \cite{cheng2015packet,diekmann2016verified,kadam2014review,zhang2015firewall}. 
For example, the aim of  \cite{cheng2015packet} is   to reduce the memory usage in  Internet traffic classification methods using the decision trees. Similarly,  in \cite{diekmann2016verified,kadam2014review,zhang2015firewall},   the authors worked on the fusion of rules  in Cisco and Palo Alto firewalls.

Some scientists tried to optimize  the rules and their sequence of appearance.
In firewalls, the rules are executed in order, and the ones appearing first have higher priority. Therefore, if a packet or data sample is  to be passed (or be blocked), performance-wise, it  is better to put the corresponding rules at the top.  In \cite{trabelsi2014dynamic,zhu2016optimization}, the authors have worked on the optimization of  rules and their order of appearance.

Due to the issue of rule preference, sometimes  conflicts are found among them. Imagine a rule lets the packet pass  while another one shall block it. In this case, the one that appear first in the list  determines the result. Many articles have focused on eliminating such inconsistencies among firewall rules \cite{al2004discovery,khummanee2013towards,maldonado2015detection,bouhoula2016security}. 

In a related effort, the authors of \cite{lorenz2015policy} presented model checking techniques and focused on finding violations of some predefined policies in the  rules. They also showed how the developed  methods can be used in  IPv6 networks.

At a bigger scale, some researchers tried to find ways of investigating whether firewalls of different parts of the same network are following similar policies or not, especially in industrial networks. For example, \cite{ranathunga2016malachite} has provided a semantic basis for expressing and comparing policies applied in\ every firewall, which in turn can  form the basis for a macro  judgment of compliance with general (e.g. organizational) policies. 

In a valuable effort, the same group  \cite{ranathunga2016case}, in addition to addressing the above problem, proposed the idea of automating firewall configuration for SCADA  systems. This idea was developed in the capacity of the ANSI/ISA 62443 standard which is intended to express  holistic security policies. The authors tried to add extended features to the ones the standard proposes in order to enable the firewall to use policies for autoconfiguration. Although this research took a step towards autoconfiguration, it was not designed to use the learned attack patterns and merely  limited a industrial network interconnections and partitioned it into zones according to the policies.

 In    \cite{adao2016localizing}, the researchers similarly worked on automating  firewall rule configuration based on a set of given policies.
In this study,  Mignis tool was used for generating the rules based on the descriptive semantics derived from policies. 
The goal was to  implement high level policies and not to learn attacks for  prevention purposes or to lower  false positives.

In \cite{ali2014firewall,liu2017firewall,chen2012first}, the authors have addressed the problem of learning firewall rules, but not by the network owner. They  investigate it from an outsider's perspective  who intends to discover the rules written in the firewall by trial and error  e.g. through sending requests and interpreting the responses.

In \cite{lar2011proactive}, a firewall is designed based on a set of  fuzzy rules. In this method, the membership level of the input packet to a set of predefined fuzzy functions is evaluated, and then the final decision  (either rejection or acceptance) is made based on the aggregated information.
The approach taken in this study is more like a fuzzy IDS whose decisions\ are enforced by an actuating IPS\ e.g.  a firewall. However, in this paper, the problem of false decisions has not yet been addressed. 

A similar research has been done in the context of http and web applications, but with learning capabilities \cite{torrano2009self}. It is assumed that the normal behavior is fed into the system in the form of XML, and any behavior that is classified as abnormal is filtered.
However, due to the lack of attention to false positives and false negatives, the performance of this method  is questionable in industrial applications.

In \cite{sairam2014implementation}, in addition to working on speed enhancements, moved  towards the  learning  capability in firewalls. They used  a Huffman tree of rules for this purpose. Learning happens by adaptations and changes in this tree.
However, the learning criterion  is merely met by crossing some hard thresholds over the measured indicators.

A network IDS based on the biological immune mechanism has been proposed in \cite{qiao2002network}. It aims lowering false positives. In this method, three monitor agents (located in different parts of the network) provide  co-stimulation signals to the intrusion detectors in order to  reduce  false positive alarms. Similarly, the authors in \cite{alshammari2007using}
used a hybrid neuro-fuzzy approach to reduce the number of false alarms in IDSs. The proposed approach was experimented with different background knowledge sets in DARPA 1999   dataset. The authors concluded through simulations that their approach required less background knowledge sets compared to other approaches. In a relevant try, \cite{joo2003neural} used neural network to build an IDS that  considers the cost ratio of false negative errors to false positive errors. The authors stated that compared with false positive errors, false negative errors incur a greater loss  thus must be lowered.  

As  explained, little work has been done in the area of automating the connection of firewalls to the units such as IDSs and attack intelligence systems for enhancing the prevention task. When we add the zero false positive requirement  to the problem, there remains no background in the literature. 
 
\section{The Proposed Approach:  Zero-False Positive Automatic  Firewall Rule Generation\label{section:proposed_model}}
In this section, we first look into extending traditional classifiers  like SVM to achieve a zero  false positive rate. By giving a set of examples, we show why this approach cannot  serve the purpose well.  Then, a novel iterative  algorithm is presented which yields zero positive results and can accommodate almost any classifier at its core. At the end of this section, we discuss how a tree-based classifier can turn the algorithm output into non-conflicting firewall rules.

%a false zero classifier as a firewall rule generator. The proposed technique is a batch mode learning for binary classification problems. In the first stage of this study, we focus on the learning technique where there is not  any feature selection on the network intrusion data set. First, main concepts of SVM classifier is briefly described. Then, we explain the proposed technique. In this study, we also take into consideration the selected features which have been determined by a real implementation study of network intrusion data and windows firewall rules.     

\subsection{Problem Description and Articulation}
%\begin{figure}
%\includegraphics[width=\textwidth, angle=0, scale=.47]{Drawing1_1.eps}
%\caption{Representation of a 2D non-separable constellation with linear %support vector machine defining the boundary.}
%\label{fig.main}
%\end{figure}
Without loss of generality, we focus on the binary classification problem. We start with the linear case. 
 Nonlinearity can be added by e.g. using kernels later though the algorithm we propose is generic and can support almost every classifier.  
%\subsection{Problem Articulation}

We assume the two classes are  nonseparable. We do not have any problem with  separable cases since they lead to zero false positive results. We do not elaborate which classifier is capable of doing so at this stage. However, for the case of linearly separable classes, there  exists  a line  that correctly separates the samples with zero error   as in e.g.  SVM~\cite{theodoridis_book}.    

To describe the research problem, we use the linear case. However, this will be relaxed later.   Fig.~\ref{fig.main}a shows a two-class set of data samples which are linearly non-separable. A linear boundary has been drawn just as an example to serve the development of subsequent formula.
Let us denote the labelled (training) samples of Class 1 (+) and Class 2 (-) by $\boldsymbol x_i ~;~i=1,...,N$.  The goal is  to find a boundary that achieves zero false positive; the statement that once decided that an input sample is +, is not  wrong.  

This means that no sample of Class 2 shall be classified as 1 (or +). You may imagine Class 1 is the set of attack samples and Class~2 is the set of normal ones in an industrial network security context.
Therefore, no
normal sample should be classified as malicious as it will be blocked and this interrupts the normal system operation.
{

Similarly, we could consider biometric security systems. In physical security systems, like fingerprint readers, false positives (false acceptances) are not tolerable, though they come at the cost of some extra false negatives. It is preferred that the granted permissions are always correct, even if sometimes legitimate users have to re-enter their fingerprints  for the reader to pick. However, in insensitive commercial applications, manufacturers tend to compromise some security for functionality as  repetitive data entry is not convenient. }

 One can achieve a zero false positive rate, but it might come at the cost of a low true negative rate (or high false negative rate). We tend to additionally minimize false negatives, or equivalently, maximize  true negatives. This could  be translated into minimizing the number of samples in Class 1 ($+$) that fall on the other side of the boundary  and are classified as ($-$).
 The two above-mentioned goals are not easy to achieve simultaneously.
In the next subsection, we mathematically show how a classic classifier works and then in  Subsection~C, extend it in  a try to make a zero false positive classifier.
\\
\subsection{ Formulation of the Classic Approach (SVM)}
There are different methods for formulation of linear classifiers facing mixed data. Normally, at some point during the development, one faces a discrete variable counting the number of misclassified samples which is usually replaced by a not-so accurate continuous  approximation to keep the problem convex \cite{theodoridis_book}.
\begin{figure}[t] %[b] %[p] puts at the end
%\begin{left}
   \centering
%   \vspace{-2ex}
    \subfloat[]{\includegraphics[width=0.5\columnwidth]{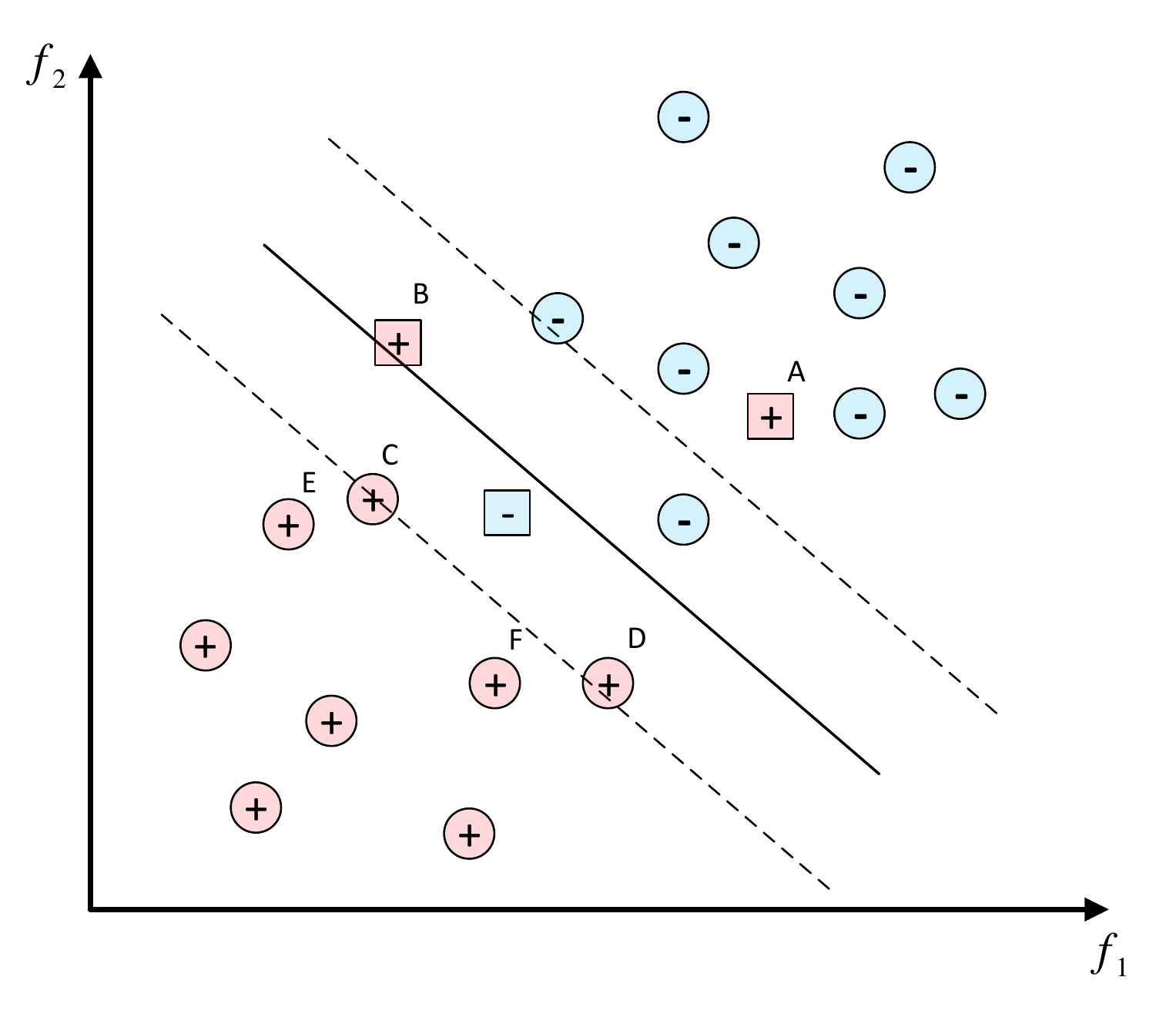}} %     \vspace{-2ex}
    \subfloat[]{\includegraphics[width=0.5\columnwidth]{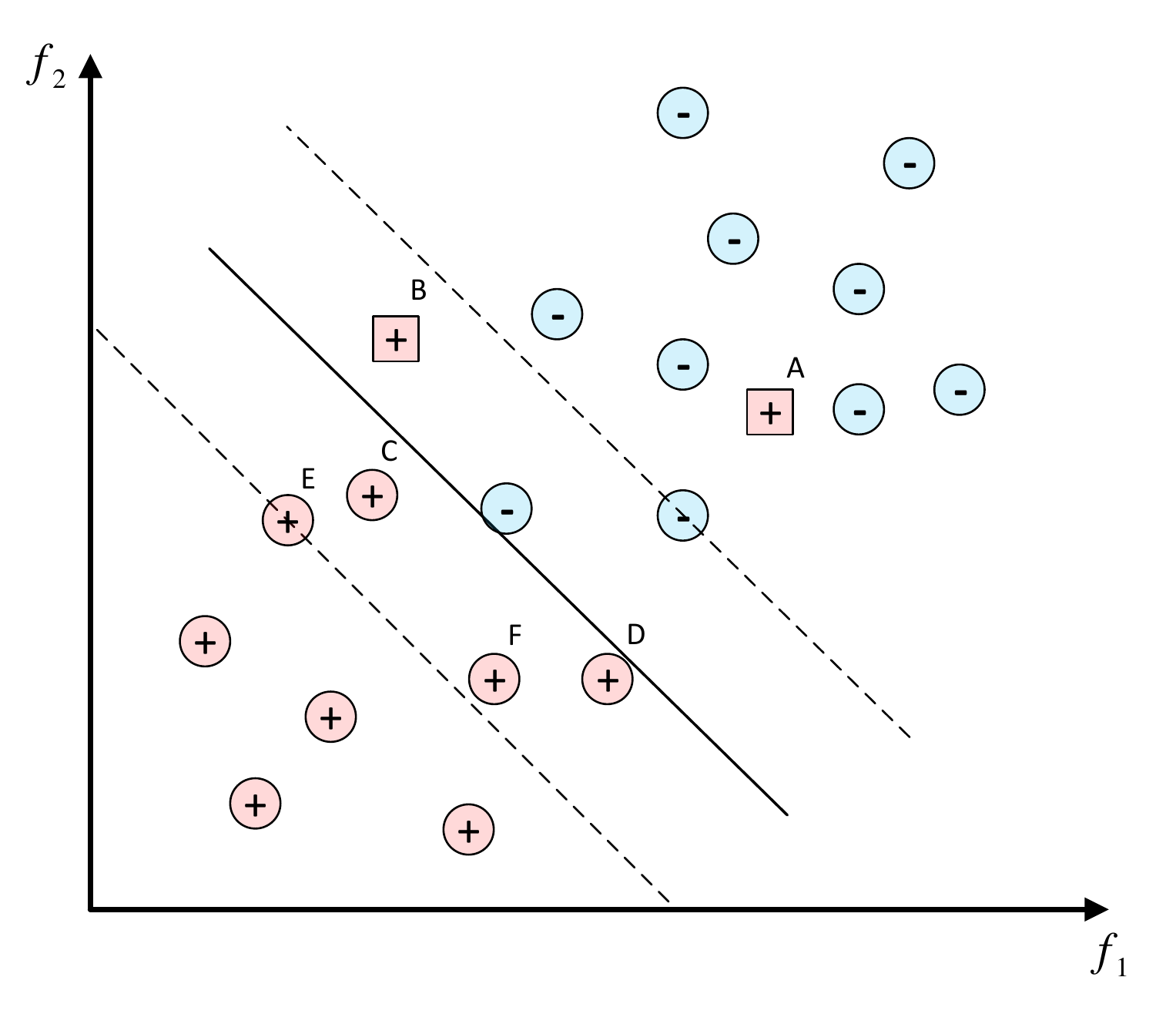}}%
\caption{(a) Traditional classification of linearly non-separable classes in a trade of between gutter  and the weight  of misclassified samples. The circles show correctly classified samples and the square ones are misclassified.
(b) Zero false-positive classification using the same classic principles but with different criteria.}
\label{fig.main}
%\end{left}
\end{figure}
We formulate the binary classification problem following the classic approach for SVM as an example. However, later during the process, we diverge from the normal definitions as we are following different targets. 
%Let us denote the training sample set by $S$ and the $i^{th}$ member of it by $\boldsymbol x_i$.
 In SVM, depending on where   the training sample $\boldsymbol x_i$  lies with respect to the linear boundary  $\boldsymbol a. \boldsymbol x_i+b_0=0$, we will have different conditions. If $y_i$ is a scalar variable whose value is +1 for  the $\boldsymbol +$ samples and -1 for the $\boldsymbol -$ ones, we can express these conditions with respect to the  boundary\ as below:

\begin{itemize}
\item 
The sample is correctly classified and falls outside the gutter (band) specified by $\boldsymbol a. \boldsymbol x_i+b_0=\pm 1$. For such samples which are shown by circles around them and lie outside the gutter in Fig.\ref{fig.main}a, we have:
\begin{align}
y_i(\boldsymbol a. \boldsymbol x_i+b_0)\geq 1
\end{align}
where equality happens when the sample (support vector) is on either of the two parallel gutter lines.
 \item The sample is correctly classified with the given $\boldsymbol a$ and $b_0$, but lies inside the gutter:
\begin{align}
0\leq y_i(\boldsymbol a. \boldsymbol x_i+b_0)< 1
\end{align}
Again, since these are classified correctly, they are shown by circles in Fig. 1(a). 
 \item The sample is  classified in the wrong class:\begin{align}
y_i(\boldsymbol a. \boldsymbol x_i+b_0)<0
\end{align}
These samples are shown by squares in Fig. 1(a).
\end{itemize}   

Some rewrite all the three in a single equation by using a slack variable as:
\begin{align}
y_{i}(\boldsymbol a.\boldsymbol x_i+b_0)\geq 1-\eta_i
\label{eq.1}
\end{align}
where  $\eta_i=0$ in the first case, $0<\eta_i\leq1$  in the second, and $\eta_i>1$ in the third. 
 Usually, the classification goal is to make the gutter as wide as possible while keeping the number of misclassified samples minimal. With the above formulation, the gutter width will be $2/||\boldsymbol a||$  \cite{theodoridis_book}. For mathematical convenience, it is usual to minimize $\frac{1}{2}||\boldsymbol a||^{2}$ instead. The cost function for such a minimization problem would be: 
\begin{align}
J(\boldsymbol a,b_0,\boldsymbol \eta)&=\frac{1}{2}||\boldsymbol a||^2+c\sum_{i=1}^{N}\mathcal{F}({\eta_i}) \label{eq.5} \\
\mathcal{F}(\eta_i)&=    
\begin{cases}
      1, & \text{if}\  \eta_i>0 \\
      0, & \text{if}\  \eta_i=0
    \end{cases}
    \label{eq.6}
\end{align}

To  remove the discontinuous functions and make the problem convex \cite{theodoridis_book},  the 
the term $\sum_{i=1}^{N}\mathcal{F}({\eta_i})$ is usually approximated by $\sum_{i=1}^{N}{\eta_i}$, which does not actually minimize the number of misclassified samples, but rather a cost function related to that: 
\begin{align}
 \underset{\boldsymbol a,b_0}{\arg\min}~ J(\boldsymbol a,b_0,\boldsymbol \eta) &=\frac{1}{2}||\boldsymbol a||^2+c\sum_{i=1}^{N}{\eta_i} \label{eq.7}
\\sbj.~to:~~&y_i(\boldsymbol a.\boldsymbol x_i+b_0)\geq 1-\eta_i \notag
\\&\eta_i\ge0 \notag
\end{align}
where the positive constant $c$ specifies the importance of the second term  compared to the gutter width.
This was a short introduction to  the classic problem formulation.

\subsection{ Extension of the Classic Approach ($\mathfrak{z}$-SVM)}

The above formulation does not guarantee a zero false positive rate. Fig.~\ref{fig.main}a shows a linearly non-separable constellation in which one sample from each class has been misclassified. Obviously, in Eq.~(\ref{eq.7}), the optimization algorithm does not have any preference over the classes. We can extend the classic formulation towards giving preference to one class and outputting zero false-positive results. If we break down the number of training samples as $N=n_{n}+n_{p}$, we can write:
\begin{align}
 \underset{\boldsymbol a,b_0}{\arg\min}~ J(\boldsymbol a,b_0,\boldsymbol \eta) &=\frac{1}{2}||\boldsymbol a||^2+c_1\sum_{i=1}^{n_{n}}{\eta_i+c_{2}\hspace{-2ex}\sum_{i=n_{n}+1}^{N}{\eta_i}} \label{eq.7}
\\sbj.~to:~~&y_i(\boldsymbol a.\boldsymbol x_i+b_0)\geq 1-\eta_i \notag
\\&\eta_i\ge0 \notag
\end{align}
where $c_1>c_2$. We refer to this classifier as $\mathfrak{z}$-SVM. The goal is to achieve something like Fig.~\ref{fig.main}b with zero false positives while keeping the false negatives as low as possible. 
%?????OPTIONAL: WE CAN WRITE THE LAGRANGIAN EQUATIONS HERE????

\subsection{The Issues with $\mathfrak{z}$-SVM}
In the previous subsection, we developed a convex formulation based on the classic approach to increase the cost of misclassification for $-$ samples. By choosing $c_1\gg c_2$, one can make the optimization algorithm try to zero the number of misclassifications for the negative samples. However, there are a number of issues with this naive approach which can make it impractical in certain scenarios.

\subsubsection{Continuous Error Approximation}

The optimization algorithm is blind to the details of the problem. In certain cases, it deceives us by manipulating the gutter width. Based on Eq.~(8), the classification error contributing to the cost function is the sum of $\eta_i$ ; $i=1,...,N$. When we want to increase $c$ (e.g. $c_1$, hoping to achieve zero false positive), the optimizer tends to make the gutter width smaller to virtually reduce the  errors (see Eq.~(\ref{eq.7})).  This is due to the fact that $\eta_i$ is a function of the gutter width itself. Fig.~\ref{fig_c_effect}a to \ref{fig_c_effect}c demonstrate this fact. 

\subsubsection{Weight Coefficients Effect} $\mathfrak{z}$-SVM  is highly sensitive to $c_1$ and $c_2$ values. It is not just their ratio that matters, their absolute values are also important along with the $||\boldsymbol a||$. 

There are three cost components in Eq.~(\ref{eq.7}). If we choose a big value for $c_1$ and a rather small one for $c_2$, we \textit{might} achieve a zero false positive rate. However, there is still a probability that this does not happen. For example, if exists a "$-$" sample deep inside the positive class region, it will contribute a large $\eta_i$ whose cost is further boosted by $c_1$. However, to make the false positive zero, the algorithm shall move the boundary line and misclassify some positive samples.  Each of these samples (let us say $j$) adds $c_2\eta_j$ to the cost. If the number of such  samples is  high, their cumulative cost might surpass\ $c_1\eta_i$. Therefore, the minimization algorithm might not output a zero false-positive result. This is highly dependent on the constellation of samples, how the classes are mixed,  and the choice of $c_1$ and $c_2$.
Fig.~\ref{fig_c_effect}d demonstrates this issue.

\begin{figure}[t] %[b] %[p] puts at the end
%\begin{left}
   \centering
      \vspace{-2ex}
    \subfloat[]{\includegraphics[width=0.5\columnwidth]{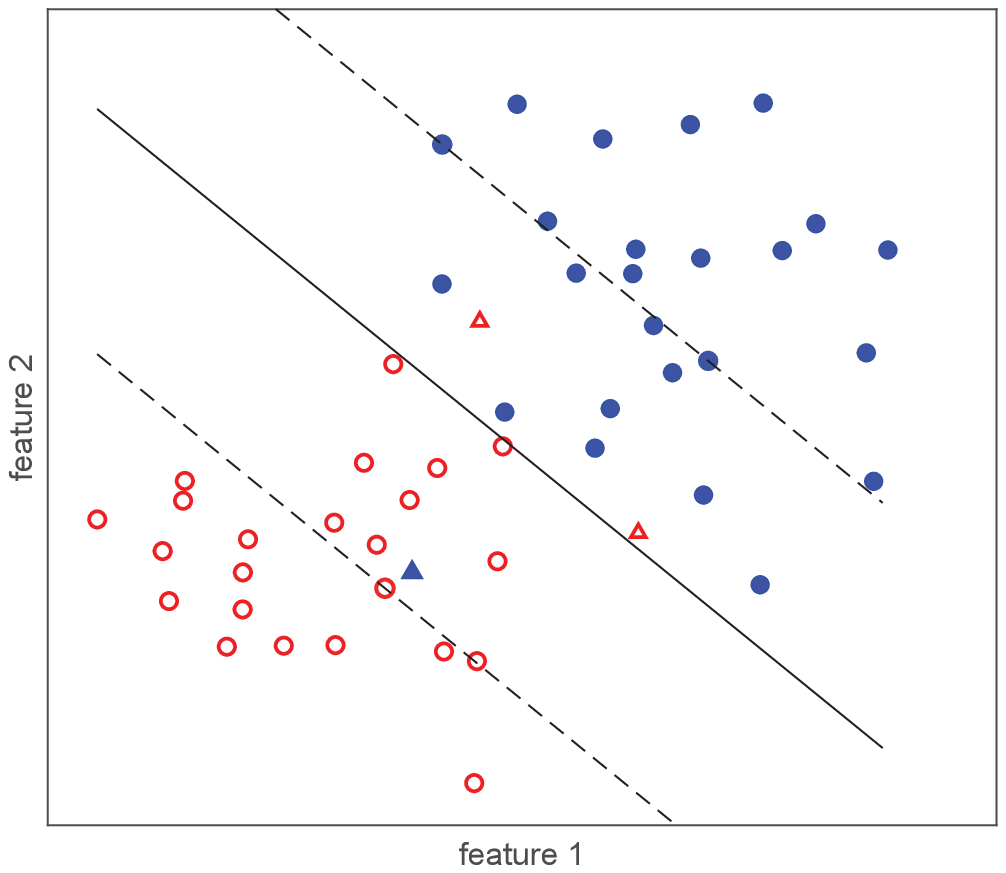}}     \subfloat[]{\includegraphics[width=0.5\columnwidth]{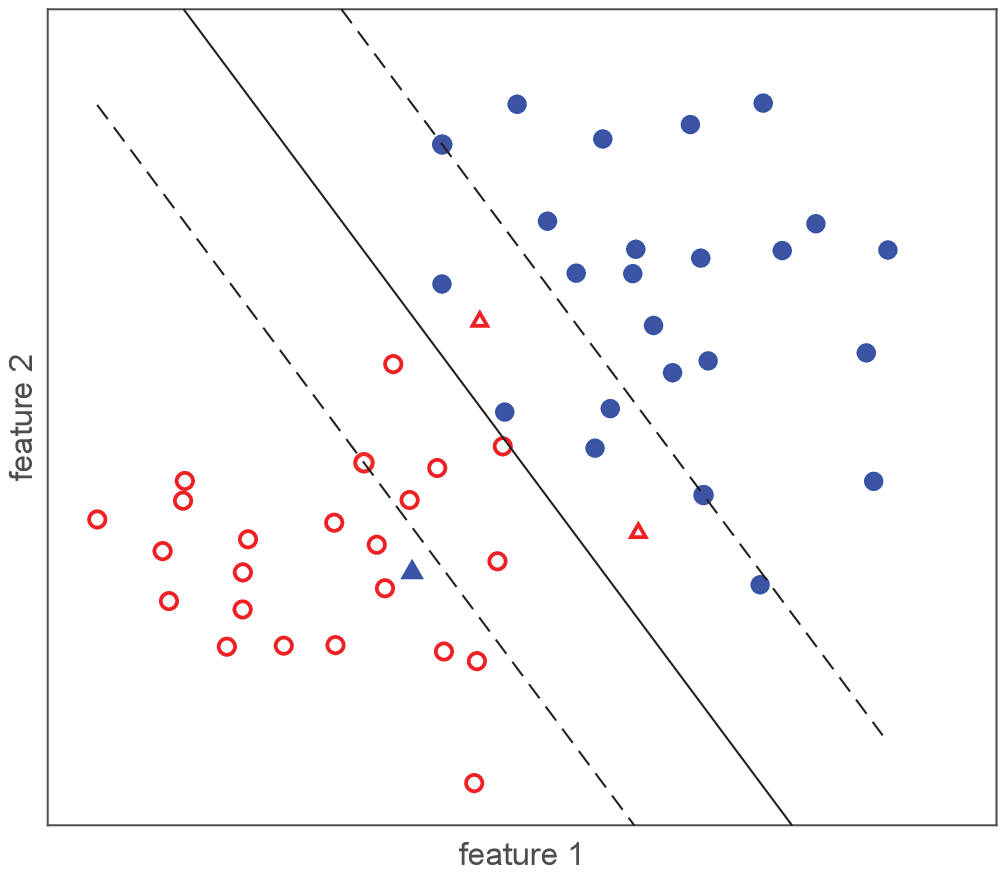}}\\     \subfloat[]{\includegraphics[width=0.5\columnwidth]{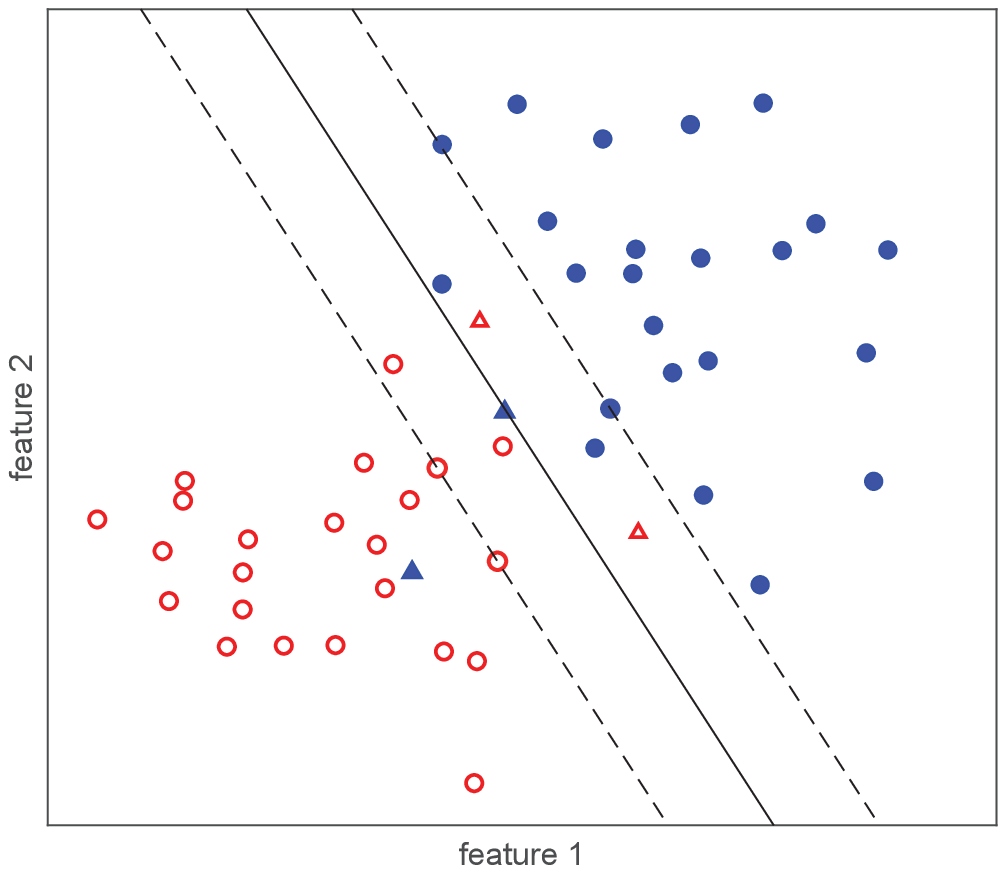}}
    \subfloat[]{\includegraphics[width=0.5\columnwidth]{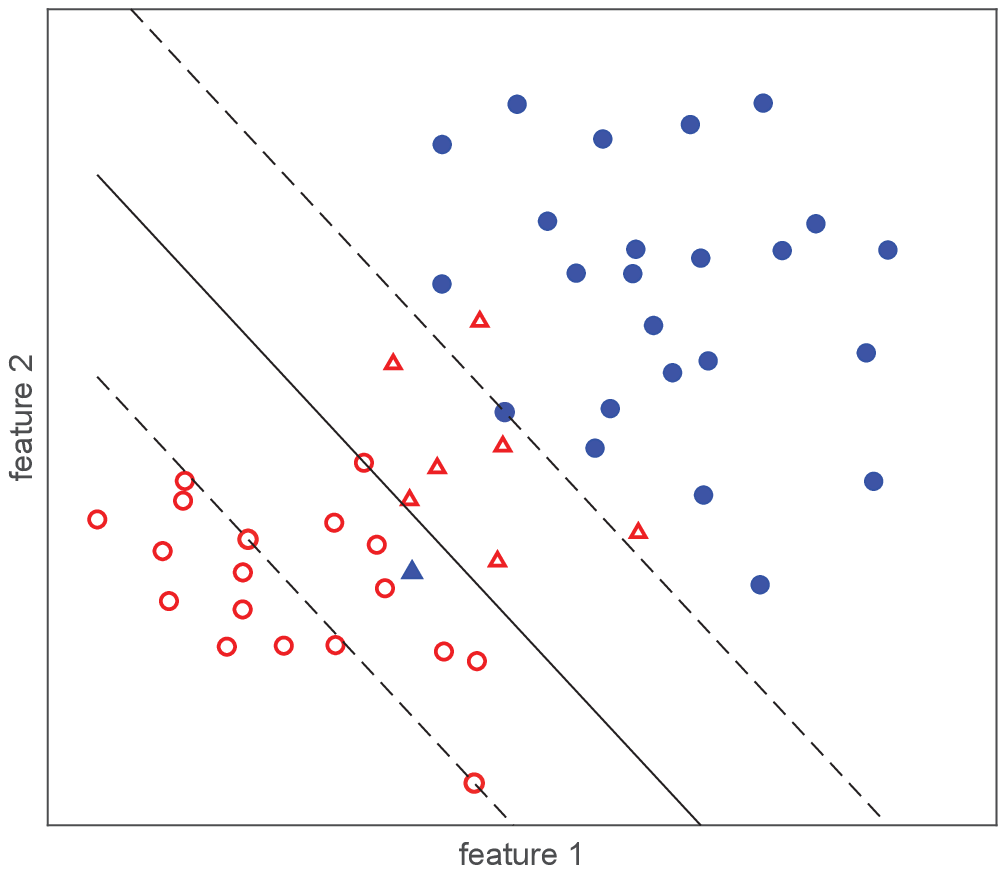}}\\ \vspace{-2ex}
     \subfloat[]{\includegraphics[width=0.5\columnwidth]{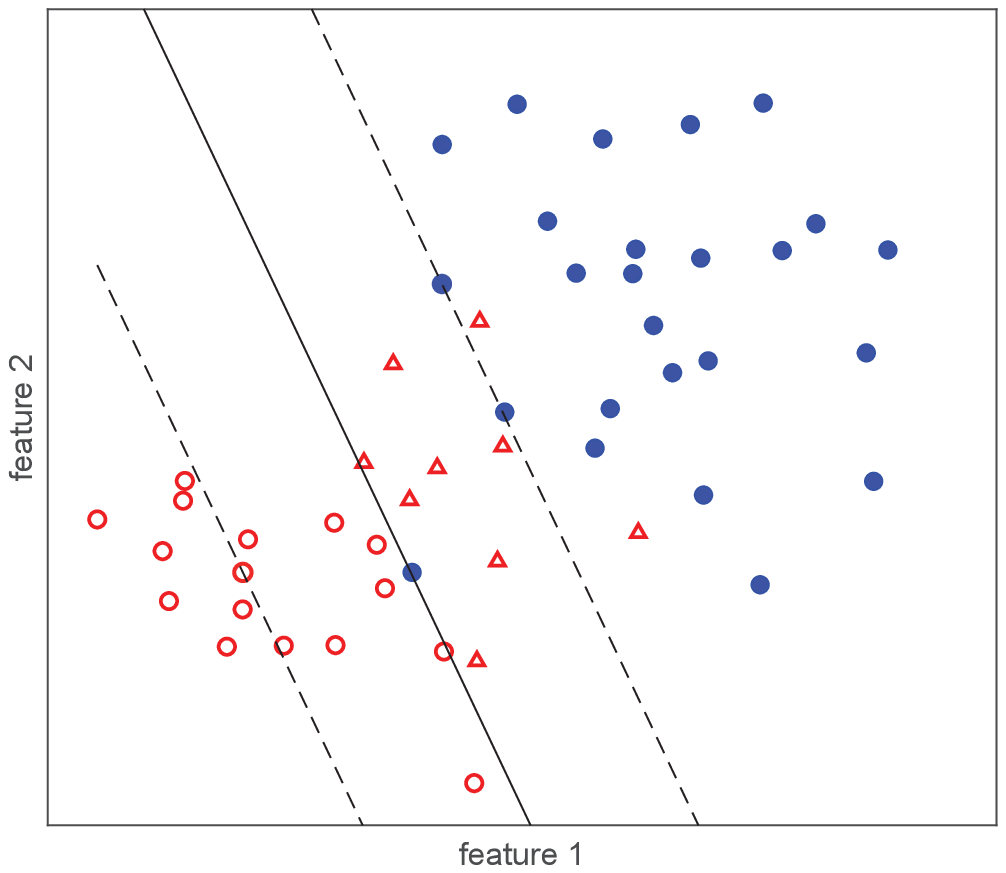}}
      \subfloat[]{\includegraphics[width=0.5\columnwidth]{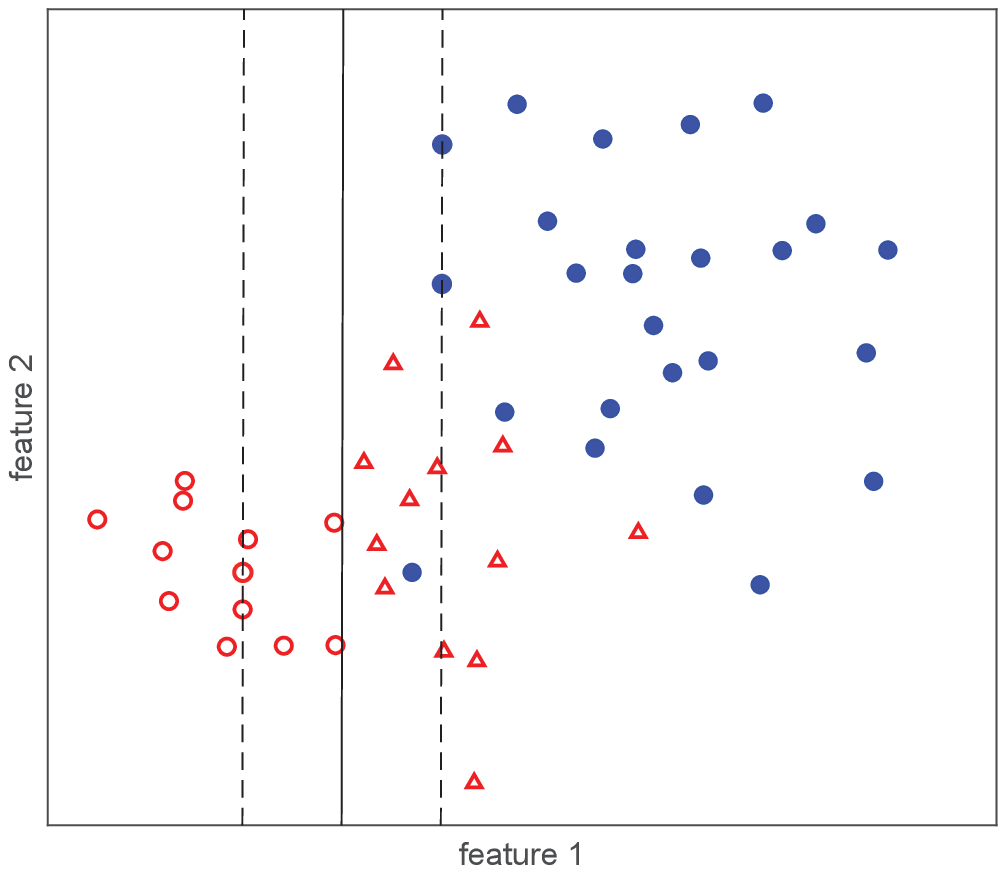}}\\
    \subfloat[]{\includegraphics[width=0.5\columnwidth]{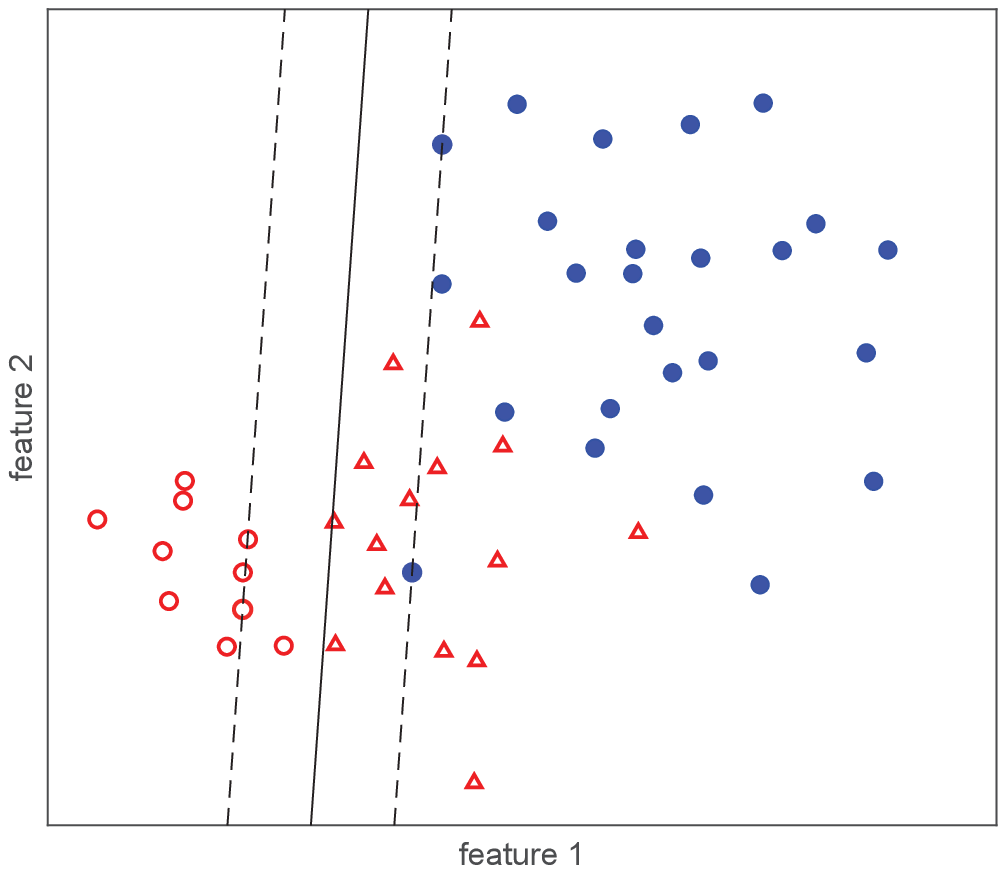}}
    \subfloat[]{\includegraphics[width=0.5\columnwidth]{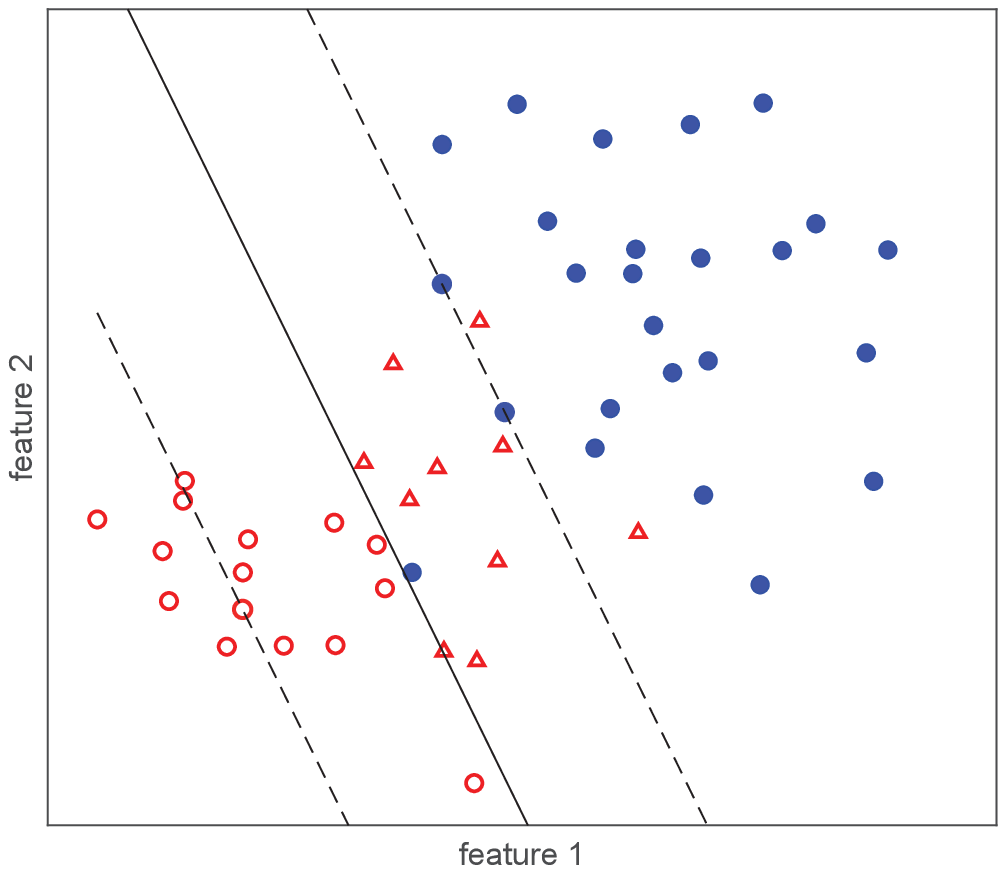}}\\ \vspace{-2ex}
        \subfloat{\includegraphics[width=1.1\columnwidth]{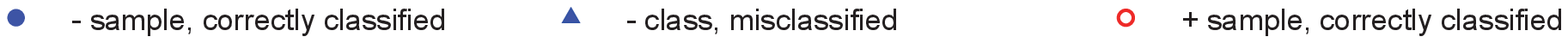}} \vspace{-2ex}
          \caption{The effect of  cost coefficient ($c$) on  $\mathfrak{z}$-SVM result: (a) $c_1=1$, $c_2=1$ (b) $c_1=5$, $c_2=5$ (c)  $c_1=50$, $c_2=50$ (d) $c_1=500$, $c_2=50$ (e) $c_1=785$, $c_2=50$ (f) $c_1=800$, $c_2=50$ (g) $c_1=1000$, $c_2=50$. (h) $c_1=80$, $c_2=5.$ }
\label{fig_c_effect}
%\end{left}
\end{figure}

A very large value for $c_1$ can solve this issue, but it induces another problem.  By choosing such a value for $c_1$, the dominant component in the cost function will be the second term. This implies that the gutter width and the $+$ samples misclassification error will have little or no effect. This might for example lead to a boundary that misclassifies too many positive sample in order to keep $c_1\sum_{i=1}^{n_{n}}\eta_i$ minimum. Fig.~\ref{fig_c_effect}e to \ref{fig_c_effect}g show this phenomenon. 
Comparison of Fig.~\ref{fig_c_effect}f and Fig.~\ref{fig_c_effect}h which have the same $c_1/c_2$ ratio  shows that the third cost component, i.e. $||\boldsymbol a||$ can significantly change the final result. Note that in the settings achieving  zero false positive rates, the number of  misclassified positive samples  are significantly different, ranging from 9 to 16, with Fig.~\ref{fig_c_effect}e having the minimum number of such samples.
The cost coefficients in Fig.~\ref{fig_c_effect}e are trivially different from those in Fig.~\ref{fig_c_effect}f, yet the results are very different. This shows how sensitive the final result is to the choice of $c_1$ and $c_2$.

\begin{figure}[b]
\center \includegraphics[width=1\columnwidth, angle =0, scale=1]{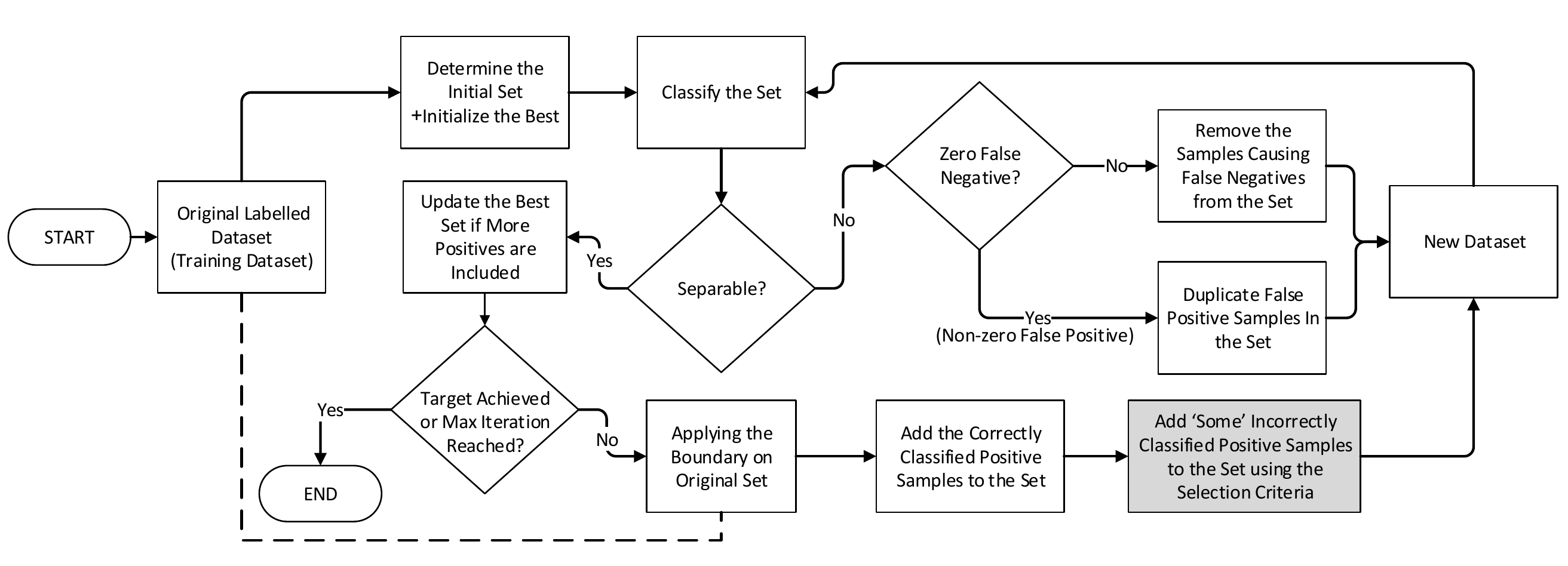}
\caption{Flowchart of  the proposed  iterative classification algorithm ($\mathfrak{z}$-Classifier). The goal is to achieve zero false positive rate.}
\label{fig.Flowchart}
\end{figure} 

\subsubsection{Sample Unbalance Problem} The training sample unbalance problem is very well known. This is not something specific to $\mathfrak{z}$-SVM. Imagine that there are merely a few negative training samples and too many positive ones. In such occasions, the cost of misclassification for the negative samples is practically capped and the optimizer might decide to minimize the cost function based on the dominant factor defined by the too-many positive samples. This implies that we need to adjust $c_1/c_2 $  per case.

The above discussions showed some of the problems of $\mathfrak{z}$-SVM; an example of traditional classifiers extension. Many of these problems pertain to the fact that the counting function (e.g. of Eq.~(\ref{eq.6})) is replaced with an approximate alternative in the convex problem development process.  One can of course stick with original optimization problems (e.g. Eq.~(\ref{eq.5})) and try to extend them  for zero false positive goal while honoring the discrete constraints (like Eq.~(\ref{eq.6})). However, the result will be a non-convex or discrete problem  at the end.

In the next section, we propose a novel algorithm for this purpose. We use classic approaches at the core of the algorithm but iteratively prune the samples based on its outputs to achieve zero false positive rates with minimum number of
false negatives. 
 Minimizing the number of false negatives will allow the firewall to capture as many malicious samples as possible and lower the load of the  sophisticated IDS, if ever installed. Finally, using a  tree classifier at the core, we test our algorithm and show how its output can be  translated into simple non-conflicting  firewall rules.

Given a classifier, false positives can be avoided if one sacrifices some of the positive samples. We suggest that this is done by removal. It can be proven that a removing strategy  always converges to a  zero-false positive result. However, the question is which positive samples 
shall be sacrificed to make this happen. We can  represent the problem as a binary satisfiability one. Imagine the positive samples are the set of variables, each being represented by a bit. A value of 0 for a bit implies that the corresponding sample shall be removed from the  set. Regarding the example of Fig.~\ref{fig.main}, this binary coding will be like $\{b_A, b_B, b_C, b_D,b_E, ...\}; b_i\in\{0,1\}$.
Now, given the classifier $\mathcal{C}$, the remaining set can be tested to see if its samples satisfy separability. Therefore, the problem is analogous to a  boolean satisfiability (SAT) problem. It can be  proven that  the problem is always satisfiable. In general, finding an  assignment of zeros and ones that has the  minimum number of variables set to one (or similarly to zero) is NP-complete, even for a satisfiable 2-SAT problem~\cite{misra2013solving}. This is sometimes called Min Ones $k$-SAT problem in the literature. The solution is not always unique though. For example, in Fig.~\ref{fig.main} and with a linear $\mathcal{C}$, both  $\{0,0,1,1,1,1,...\}$ and  $\{0,1,1,0,1,1,...\}$ will satisfy the separability thus the resultant boundaries can be acceptable. Although there could be several answers, some additional constraints in the cost function (like maximum gutter width or distance to the border) can reduce the  set of answers.   

\subsection{The Proposed  Zero-FP Classifier ($\mathfrak{z}$-Classifier)\label{sec:proposed}}

%\begin{figure*}[t] %[b] %[p] puts at the end
%   \centering
%    \subfloat[]{\includegraphics[width=0.5\columnwidth, height=0.45\columnwidth]{svm_evolution1.png}} %    \subfloat[]{\includegraphics[width=0.5\columnwidth, height=0.45\columnwidth]{svm_evolution2.png}} %    \subfloat[]{\includegraphics[width=0.5\columnwidth, height=0.45\columnwidth]{svm_evolution3.png}}
%    \subfloat[]{\includegraphics[width=0.5\columnwidth, height=0.45\columnwidth]{svm_evolution100.png}}\\ %\vspace{-2ex}
%        \subfloat{\includegraphics[width=2\columnwidth]{legend.eps}} \vspace{-2ex}
%\caption{Demonstration of  zero false positive classification using the proposed   technique for one particle with SVM as $\mathcal{C}$. The goal is to find a  boundary that does not misclassify the negative samples  (left/blue). Bold samples indicate inclusion   in the particle's set. (a) iteration No. 1 (b) No. 2 (c) No. 3 (d) No. 100.}
%\label{fig.samplex}
%\end{figure*}
The proposed algorithm is an iterative classification method which achieves a zero false detection rate for the positive class.
It is a swarm algorithm whose particles are reduced datasets. Each reduced dataset consists of all the negative samples and some of the positive ones. In this sense, we can represent each particle with a binary coding similar to before i.e. $\{b_1, b_2,  ..., b_{n_{p}}\}; b_i\in\{0,1\}$, where for the sake of simplicity, we have removed the 1s of the negative class members as they are always present.
Each particle follows an algorithm which is illustrated in Fig.~\ref{fig.Flowchart}.
The grey box is where it uses the knowledge obtained by other particles.

Each particle starts with an initial dataset which is derived from the original one by removing some of the positive samples in order to become separable. Ideally, this initial dataset has only one positive sample, and all the negative ones. Given the classifier  $\mathcal{C}$, one can do an initial classification on the original dataset and use the true positives as candidates to make this initial dataset. 
Every particle's reduced dataset goes through a classification by using $\mathcal{C}$. In each iteration, if the result shows separability of the  set (zero false positive and zero false negative), the set is compared against the best result obtained by other particles so far. The comparison criterion could be a predefined fitness function e.g. the number of positive samples included as we intend to minimize the number of false negatives. If the particle' reduced set yields  a better score, the best record is replaced with its set. Then, the
boundary model obtained through the process is applied to the original dataset (as a test set) and the new true positives are tentatively added to the particle's set for the next iteration. This is done by setting their flag to 1. Moreover, some extra positive samples that are misclassified by this boundary are also selected and added to the set.
There can be different selection criteria for this purpose, however, we suggest that one uses the global best (as in Particle Swarm Optimization) for this purpose. For example, one can use weighted random selection to pick $k$ samples from the set of falsely classified positive samples (of the original set) and add them to the particle's set. Obviously, the positive samples included in the best set should have more weight in this selection. 

On the other hand, throughout the iterations, if a particle's set becomes inseparable at some point, it will undergo a pruning operation. The pruning is initially done by removing the positive samples that have created false negatives. If this does not help (which will be known in the next iteration(s)) or the source of inseparability is non-zero false positive, the weight of the negative samples contributing to the false positives is increased. This can be done by e.g. duplicating those negative samples in the particle's dataset. As it can be seen, the  reduced dataset of particle is not necessarily a subset of the original dataset yet they are  related. The pruning  continues over  iterations until the set becomes separable again. The whole addition/removal process terminates when either the target false negative is met or the maximum number of allowed iterations is reached. Algorithm~1 shows $\mathfrak{z}$-Classifier.   

\begin{algorithm} [t]
   \caption{The Proposed Iterative  Classification Technique  with Zero False Positive  ($\mathfrak{z}$-Classifier)}
   \label{algor}
    \begin{algorithmic}[1]
        \REQUIRE  Training Set($\mathcal{T}$), Performance Target/Stop Criterion
        \ENSURE  ~Zero False Positive Classification  
%\hline  \NoNumber ~
        \STATE Create initial separable particle sets  $S_1,...S_k$
        \STATE Determine the $Best$ set (with the most positive samples)
\WHILE {Stop Criterion/Performance Target is not met}
               \FOR {each $S_i$}
               \STATE Classify $S_i$ with $\mathcal{C}$ and find the boundary $\mathcal{B}_i$
                       \IF {$S_i$ is separable}
                        \STATE Update the $Best$     if $S_i$ is more fit                         \STATE Apply $\mathcal{B}_i$ to $\mathcal{T}$ \& add the  true  positives  to $S_i$
\STATE Use the $Best$ and $S_i$ knowledge to add $k_{i }$ false positive samples to $S_i$                 
                       \ELSE
                       \IF {there is false negatives}
                         \STATE Remove false negative samples from $S_i$
                       \ELSE
                               \STATE Duplicate false positive samples in $S_i$
                       \ENDIF
         \ENDIF
               \ENDFOR
               \ENDWHILE
\end{algorithmic}
\end{algorithm}

%Fig.~\ref{fig.samplex} depicts the first three steps as well as the final step of the proposed algorithm applied to a 2D synthetic data set with 7500 samples. In this specific case, SVM has been used as the core classifier. The algorithm output can be directly translated into firewall rules though. In that case, a tree-based classifier is preferred. This is explained  in the next section.
%\vspace{-2ex}
\section{Evaluation Results and Discussions\label{section:results}}
%\begin{figure*}[!t] %[b] %[p] puts at the end
%   \centering
%    \includegraphics[width=2\columnwidth, height=0.6\columnwidth]{tree.eps}
%    \caption{Demonstration the  tree output of $\mathfrak{z}$-Classifier % at the end of the 10th iteration for KDD CUP'99 dataset. +1 = malicious class  -1 = normal class.}
     % \vspace{-2ex}
%\label{fig.kdd}
%\end{figure*}
Here, we  evaluate $\mathfrak{z}$-Classifier
by testing it on  two  datasets. We take the CART algorithm  as our classifier. Classic firewalls can  define rules by combining  clauses that put constraints on the values  of features. However, each clause merely constrains one feature. For such construction, tree-based classifiers like CART produce more compatible results.  In the  simulations, the population size was  5 and {$k_{i}$ was   increased exponentially by a factor of 1.5 (from 1) for faster convergence. It was reset to  1 every time the set became non-separable. }  
\subsection{Power System (Smart Grid) Dataset}

Smart grids or power systems  have supervisory control systems interacting with different smart electronic devices. They   are complemented by network monitoring devices such as SNORT and Syslog. Some researchers created a dataset of attacks launched in power systems  \cite{pan2015developing,hink2014machine}. In their scenario, it is assumed that an actor gains  access to a substation network and poses an insider threat by issuing commands from the substation switch. The  scenarios studied were (1) Short-circuit fault, (2) Line maintenance (3) Remote tripping command injection (Attack), (4) Relay setting change (Attack), and (5) Data Injection (Attack) \cite{pan2015developing,hink2014machine}. This study worked on events categorized in binary classes which contained 37 event scenarios grouped as either  attack (28 events) or normal operations (9 events). We used a subset of the dataset with 1861 normal and 3415 attack samples described by 128 features. Date and time  were taken out of the dataset records as we did not want the classifier to rely on those for detection. The results of applying $\mathfrak{z}$-Classifier on this dataset are  shown in   Table \ref{tab.confusion2}. As one can see and similar to before, the proposed classifier always maintains a zero false positive output.

\begin{table}[b]
\caption{$\mathfrak{z}$-Classifier performance  on the power  grid dataset  of \cite{pan2015developing,hink2014machine}}
\begin{center}
 \begin{tabular}{||c c c c c||} 
 \hline
Iterations  & ~~~TN~~~ & ~~~TP~~~ & ~~~FN~~~ & ~~~FP~~~\\ [0.5ex] 
 \hline\hline
 10 & 1861 & 320 & 3095 & 0 \\ 
 \hline
 50 & 1861 & 888 & 2527 & 0\\
 \hline
 100 & 1861 & 2211 & 1204 &0\\
 \hline
 500 & 1861 & 3231 & 184&0 \\
 \hline
 1000 & 1861 & 3353 & 62 &0\\ [1ex] 
 \hline
\end{tabular}
\end{center}
\label{tab.confusion2}
\end{table}

\subsection{Results of KDD  CUP'99 Dataset}
KDD Cup'99  dataset  has been  developed by MIT Lincoln Laboratory in 1999 \cite{KDDCup}. %\cite{KDDCup}. 
We  use a subset of the dataset that has 10\% of the total records (494,021). Every record  has 41 features and contains a label for the sample class. Along with normal samples, there are 24 types of   attacks in the training set which fall into  four categories. However,  we turned all the attack labels  into malicious so that the algorithm is left with a decision of ACCEPT or REJECT for each sample. 
%The resultant decision tree  is depicted in Fig.~\ref{fig.kdd}. Since the complete tree was too big to be plotted, we have reported the output after 10 iterations. Even this simple classifier can capture 138,869 out of 204,458 malicious samples with zero false positive. 
Even a simple classifier obtained after 10 iterations can capture 138,869 out of 204,458 malicious samples with zero false positive. 
The complete tree after 1000 iterations  yields much better results which are reported in   Table \ref{tab.confusion1}.
Please note that just like any learning classifier, $\mathfrak{z}$-Classifier is also prune to over-fitting. We noticed over-fittings after the 100th iteration with KDD CUP'99 dataset. One can adopt traditional best practices to  avoid this phenomenon. 
\begin{table}[t]
\caption{$\mathfrak{z}$-Classifier performance  on KDD CUP'99 dataset.}%\cite{KDDCup}.}
\begin{center}
 \begin{tabular}{||c c c c c||} 
 \hline
Iterations  & ~~~TN~~~ & ~~~TP~~~ & ~~~FN~~~ & ~~~FP~~~\\ [0.5ex] 
 \hline\hline
 10 & 289,542 &  138,869 & 65,589 & 0 \\ 
 \hline
 50 & 289,542 & 204,167 & 291 & 0\\
 \hline
 100 & 289,542 & 204,425 & 33 &0\\
 \hline
 500 & 289,542 & 204,432 & 26&0 \\
 \hline
 1000 & 289,542 & 204,432 & 26 &0\\ [1ex] 
 \hline
\end{tabular}
\end{center}
\label{tab.confusion1}
\end{table}

Translation of the $\mathfrak{z}$-Classifier output to firewall rules is a trivial task now.  For the  tree described before, the first few rules of a default-allow firewall will be like:

\begin{flushleft}IF $x_{29}<0.255$ \& $x_{35}\geq0.145$ \& $x_{32}<97.5$ then  REJECT
IF $x_{29}\geq0.255$ \& $x_{37}\geq0.055$ \& $x_{23}\geq4.50$ then REJECT\\~$\vdots$\end{flushleft}%\vspace{-2ex}
in which $x_{29}$ ('same\_srv\_rate') shows  the percentage of connections to the same service, $x_{35}$ ('dst\_host\_diff\_srv\_rate') is the percentage of connections to different services, $x_{32}$ ('dst\_host\_count') is the number of connections to the same destination host, $x_{37}$ ('dst\_host\_srv\_diff\_host\_rate') shows the percentage of connections to the same service coming from different hosts, and $x_{23}$ is the number of connections to the same host as current connection in the past two seconds.% \cite{KDDCup}. 

\section{Conclusion\label{section:conclusion}}

 Rejecting legitimate traffic by mistake is not tolerable in  real-time industrial systems e.g. power systems. Intrusion detection systems, if ever installed, also have limited capacity of inspection and it is preferred that the attacks are filtered at intrusion preventions systems such as firewalls as much as possible. 
In this study, a new classifier is introduced that can help in building self-organizing learning firewalls.  
Current classifiers tend to minimize a generic cost function which does not necessarily yield zero false positive results. In this paper, we proposed an  algorithm that can turn any generic classifier into a zero false positive one. One can use this classifier  to build a self-organizing and learning firewalls.  
In the design of this new approach  zero  false positive (or negative) rate has been set as the goal, while the false negative (or positive) rate is kept minimum. The designed classifier was tested on two datasets; a power grid dataset and KDD CUP'99 dataset.  We also showed how a tree-based  classifier can help us automate writing  zero-false positive firewall rules.

\appendices
\section{Satisfiability with the  Removal Strategy\label{appendix2}}

\textit{\textbf{Theorem 1.}  }Given a dataset with two classes and a sound classifier like $\mathcal{C }$, an iterative removal strategy that takes the false positive samples out in each iteration will always lead to a zero false positive result. \\

\textit{Proof}. 
Assume that the training dataset is $(x_i,y_i)$ where $i=1,2,...,m$ and $y_i\in\left\{ 1,-1 \right\}$. The zero FP algorithm applies a sound classifier $\mathcal{C}$  as a core classifier which minimizes a predefined cost function like $J$. 

First, the algorithm starts on the complete and original dataset. After classification, there are two types of misclassified samples which make the classification errors. The first one corresponds to false negatives $(E^{P\rightarrow N})$, and the second one is related to false positives $(E^{P\rightarrow N})$. The cost function of the classifier is $J(E_{0}^{P\rightarrow N}, E_0^{N\rightarrow P})$, where the 0 superscript shows the iteration number. According to the zero FP algorithm, the positive samples which make $E^{P\rightarrow N}$    must be removed from the original data set at this stage. After applying the classifier boundary to the new data set (after removal), $E^{P\rightarrow N}$ will be zeroed. Hence, the cost function of the zero FP algorithm at this stage becomes $J(E_1^{N\rightarrow P})<J(E_0^{P\rightarrow N},E_0^{N\rightarrow P})$. This inequality can be rewritten as:
\begin{align}
J(E_1^{N\rightarrow P})=r_1~J(E_0^{P\rightarrow N},E_0^{N\rightarrow P})
\label{eq.9}
\end{align}
where $0\leq r_{1}\leq1$.  
Sequentially, the sound classifier provides a new classification boundary for the modified dataset whose cost function is $J(E_{2}^{P\rightarrow N}$, $E_2^{N\rightarrow P})$. Since the dataset is modified,  \(\mathcal{C}\)    tried to minimize the misclassification errors so that it is less than (or in the worst case equal to)  the previous iteration, i.e. $J(E_2^{P\rightarrow N}$ , $E_2^{N\rightarrow P})< J(E_1^{N\rightarrow P})$. Hence,
\begin{align}
J(E_{2}^{P\rightarrow N} ,E_2^{N\rightarrow P})=q_1~J(E_1^{N\rightarrow P})
\label{eq.10}
\end{align}
where $0\leq q_1\leq1$. Similarly the algorithm removes the positive samples which have been misclassified. At the $k^{th}$ iteration, the cost function will be:      
\begin{align}
J(E_{k}^{P\rightarrow N} ,E_{k}^{N\rightarrow P})=q _{k}~J(E_{k-1}^{N\rightarrow P}); ~~~0\leq q_{k}\leq1
\label{eq.13}
\end{align}
and at the $k+1^{th}$ one,
\begin{align}
J(E_{k+1}^{N\rightarrow P})=r_{k}~J(E_{k}^{P\rightarrow N},E_{k}^{N\rightarrow P}); ~~~0\leq r_{k}\leq1
\label{eq.14}
\end{align}
This way, at each iteration a reduced set is prepared for  the next iteration. By substituting Eq.~(\ref{eq.13}) in Eq.~(\ref{eq.14}), 
\begin{align}
J(E_{k+1}^{N\rightarrow P})=(r_{k}~q _{k})~J(E_{k-1}^{N\rightarrow P})
\label{eq.15}
\end{align}
Similarly, 
\begin{align}
J(E_{k+1}^{N\rightarrow P})=\left(\Pi_{k=1}^{N}(r_{k}~q _{k})\right)~J(E_0^{P\rightarrow N},E_0^{N\rightarrow P})
\label{eq.16}
\end{align}
As $\Pi_{k=1}^{N}(r_{k}~q _{k})$ goes to zero, we have 
\begin{align}
\lim _{k\rightarrow\infty}J_{}(E_{k+1}^{N\rightarrow P})=0
\label{eq.17}
\end{align}
Notice that $\text{max}(N)$ is the number of positive samples. 
Therefore, the cost function of the zero FP algorithm with a sound cost-minimizing classifier will converge to zero, and all negative samples will be truly classified.

%
%\section*{Acknowledgment} This research was in part supported by  Institute for Research in Fundamental Sciences (IPM) with No. CS1397-4-77. 

% by themselves when using endfloat and the captionsoff option.
\ifCLASSOPTIONcaptionsoff
  \newpage
\fi
% trigger a \newpage just before the given reference
% number - used to balance the columns on the last page
% adjust value as needed - may need to be readjusted if
% the document is modified later
%\IEEEtriggeratref{8}
% The "triggered" command can be changed if desired:
%\IEEEtriggercmd{\enlargethispage{-5in}}

% references section
% can use a bibliography generated by BibTeX as a .bbl file
% BibTeX documentation can be easily obtained at:
% http://www.ctan.org/tex-archive/biblio/bibtex/contrib/doc/
% The IEEEtran BibTeX style support page is at:
% http://www.michaelshell.org/tex/ieeetran/bibtex/
%\bibliographystyle{IEEEtran}
% argument is your BibTeX string definitions and bibliography database(s)
%\bibliographystyle{IEEEtranS}
%\bibliography{referencesE}
\bibliographystyle{IEEEtran}
% ument is your BibTeX string definitions and bibliography database(s)
%\bibliographystyle{IEEEtranS}
%
\bibliography{refers}
%
%\printindex
\end{document}